 \documentstyle[12pt]{article} 
\begin{document} 

 \begin{center} {\Large \bf 
  Reply Comment: \\
  Comparison of Approaches \\
  to Classical Signature Change  
 \\[1cm]} 

  {\large  Charles Hellaby \\} 
  {\small \it Department of Applied Mathematics, University of Cape 
Town, Rondebosch, 7700, South Africa \\} 
  E-mail: {\tt cwh@maths.uct.ac.za} \\[4mm] 

 {\large \&} \\[4mm] 

 {\large Tevian Dray} \\ 
 {\small 
 Department of Mathematics, 
 Oregon State University, 
 Corvallis, 
 Oregon 97331, 
 USA \\ 
 School of Physics and Chemistry,
 Lancaster University, 
 Lancaster,
 LA1 4YB, 
 U.K.} \\ 
 E-mail: {\tt tevian@math.orst.edu} 
 \\[1cm] 

 Phys. Rev. D52, 7333-7339 (1995).\\
 gr-qc/9601040 \\[1cm]


 \end{center} 

 \noindent{\large \bf ABSTRACT} 

     We contrast the two approaches to ``classical" signature change 
used by Hayward with the one used by us (Hellaby and Dray).  There is 
(as yet) no rigorous derivation of appropriate distributional field 
equations.  Hayward's distributional approach is based on a postulated 
modified form of the field equations.  We make an alternative 
postulate.  We point out an important difference between two possible 
philosophies of signature change --- ours is strictly classical, while 
Hayward's Lagrangian approach adopts what amounts to an imaginary 
proper ``time" on one side of the signature change, as is explicitly 
done in quantum cosmology.  We also explain why we chose to use the 
Darmois-Israel type junction conditions, rather than the Lichnerowicz 
type junction conditions favoured by Hayward.  We show that the 
difference in results is entirely explained by the difference in 
philosophy (imaginary versus real Euclidean ``time"), and not by the 
difference in approach to junction conditions (Lichnerowicz with 
specific coordinates versus Darmois with general coordinates). 

 \newpage

 \noindent{\large \bf DIFFERENCES IN APPROACH} 
 \\[-2mm] 

 \noindent{\it Meaning of `classical'} 

     The most fundamental difference between the approach of Hayward 
and that of Hellaby and Dray (us) is in the treatment of the region of 
Euclidean signature.  We (Hellaby and Dray) [1,2] adopt a strictly 
classical view of this region, as do Ellis et al [3-7] and Dray, 
Manogue and Tucker [8-11], whereas Hayward [12-14], in his Lagrangian 
approach [12], works with squared momenta that change sign, which is 
equivalent to retaining the imaginary proper ``time" and momentum 
fields that appear in the quantum cosmological treatment [see e.g. 
17,18], obtaining a classical version of the quantum cosmological 
result.  
 In an interesting comparison, Kerner and Martin [20-21] have done a 
quantum cosmological version of the strictly classical signature 
change, retaining the real ``time" of the classical treatment of 
Euclidean signature, so that the wave function is oscillatory in the 
Euclidean region as well as the Lorentzian region.  It is not 
surprising that these four approaches (classical or quantum cosmology 
with real or imaginary Euclidean ``time") all lead to somewhat 
different results.  None of them can claim to have any experimental or 
observational support. 
 \\[-2mm] 

 \noindent{\it Appropriate form of the EFEs} 

     The Einstein Field equations were derived assuming a 
 constant signature metric.  
 For signature change, the most basic question to be addressed is that 
of the applicable form of the Einstein equations at a change of 
signature.  Various postulates have been made and will be discussed 
below.  The differences between them can be reduced to the question of 
whether the Euclidean proper ``time" is real or imaginary.  The 
applicable form of the conservation equations is similarly open to 
question.  
 \\[-2mm] 

 \noindent{\it Distributional EFEs } 

     Formal calculation of the Einstein tensor from a step function 
metric results in products of distributions, so another approach is 
required to obtain a well defined distributional form of the Einstein 
field equations (EFEs).  Hayward has shown [12,14] that the usual form 
of the EFEs for a lapse function metric may be interpreted 
distributionally, but the derivation of these equations requires 
certain (unstated) smoothness assumptions.  This does not in itself 
support those assumptions or that form of the distributional EFEs.  We 
postulate an alternative form for the distributional EFEs based on 
more physical (and less restrictive) assumptions.  

     The approach of Kossowski and Kriele [15-16] also makes 
restrictive smoothness assumptions.  Their comparison of two 
``classical'' approaches investigates under what conditions they are 
equivalent, showing that in general they are not; and in each case 
under what conditions the Einstein tensor is bounded and/or 
continuous.  However, they require the full  4-dimensional matter 
tensor to be not only bounded but also smooth through $\Sigma$.  This 
means their proposal (a) is not what is done by us and by Ellis et al 
[3], so that their remark 2 criticising [3] and their very restrictive 
results from theorems 9 and 10 are not applicable. 
 \\[-2mm] 

 \noindent{\it Choice of junction conditions} 

     The standard view for realistic Lorentzian solutions of 
Einstein's field equations is that the metric should be $C^1$ (on one 
or more 3-surfaces) and piecewise $C^3$ (everywhere else).  The 
surface of the moon is an example of a timelike surface dividing 
regions of different density.  The time of decoupling in the early 
universe is an example of a spacelike surface dividing regions of 
different pressure.  In other words, some components of the 
Einstein/matter tensor have jumps at junctions.  How should a surface 
of signature change be treated?  It is entirely reasonable to permit 
similar or geater discontinuities at a signature change --- a change 
which is far more fundamental than a change in the equation of state. 

     In the Lagrangian approach [12], the arguments for the vanishing 
of the momenta are similar to those used in quantum cosmology [17,18] 
for the vanishing of the extrinsic curvature.  To satisfy the Einstein 
field equations (EFEs) and conservation equations, {\it through} the 
signature change surface $\Sigma$, implicitly requires the metric and 
the matter fields (in this case a scalar field $\phi$) to be $C^3$.  
In contrast, the Lichnerowicz junction conditions for the case of no 
signature change (which are equivalent to the constant signature 
Darmois junction conditions) say that if you can find admissible 
coordinates spanning $\Sigma$ in which the 4-metric is $C^1$, then you 
have a good matching.  At a signature change, this continuity (whether 
$C^1$ or $C^3$) also requires the choice of a particular class of 
coordinates 
 --- lapse-shift coordinates with a squared lapse function, such as 
$N(t) = t$.  We point out below that this unusually high degree of 
continuity can only be achieved in a limited sense, owing to the fact 
that the metric is degenerate at $\Sigma$, meaning the inverse metric 
is divergent. 

     The alternative choice of the Darmois junction conditions is 
attractive because they are expressed in terms of invariants of the 
signature change surface $\Sigma$ 
 --- the intrinsic (3-d) metric ${}^3\!\!g_{ij}$ and the extrinsic 
curvature $K_{ij}$.  In the familiar constant signature case, if the 
 one-sided limits of these two on one side of $\Sigma$ (i.e. in 
manifold with boundary $M^+$) equal the 
 one-sided limits on the other side ($M^-$), then there is a good 
matching, and it is possible to find coordinates spanning $\Sigma$ 
which make the metric and the inverse metric at least $C^1$.  The 
 (one-sided limits of these) invariants do not depend on the choice of 
coordinates in the enveloping manifolds $M^-$ and $M^+$.  In 
particular, for a spacelike surface, they do not depend on the choice 
of time or lapse function.  For the case of a change of signature at 
$\Sigma$, careful analysis [1] showed that no modification of these 
conditions, or any special coordinate choice is required.  The Darmois 
conditions are blind to the change in signature, just as the EFEs are 
blind to the metric signature.  Of course it is no longer possible to 
find coordinates that make both the metric and its inverse $C^1$, but, 
as above, the covariant metric can be made arbitrarily smooth.  Thus 
all the main results in [1] are coordinate invariant.  Supplementary 
junction conditions on the equation of state or the matter fields may 
also be imposed, and such conditions were given in [1] for an 
electromagnetic field, as well as the consequences of the EFEs at a 
signature change in an arbitrary electrovac model, in equations 
 (78)-(80). 

     Standard conservation can be shown to hold as a consequence of 
the constant signature Darmois (or Lichnerowicz) conditions.  At a 
signature change, the same conditions lead to extra jumps in physical 
quantities, such as the density, and a surface term in the derived 
conservation law.  Conservation is not abandoned, but it has to be 
modified.  Stronger conditions than those in [12] are required to 
restore the conservation law to its familiar form, for the strictly 
classical case. 
 \\[3mm] 

 \noindent{\large \bf THE CASE OF A HOMOGENEOUS SCALAR FIELD} 
 \\[-2mm] 

 \noindent{\it The Field Equations for a squared-lapse metric} 

     Consider the case of a homogenous isotropic universe with scale 
factor $a(t)$ and squared lapse function $N(t)$, filled with a scalar 
field $\phi(t)$ with arbitrary potential $V(\phi)$.  The line element 
 \begin{equation}   ds^2 = -N(t) dt^2 + {}^3\!\!g_{ij} dy^i dy^j = 
   -N(t) dt^2 + a^2(t) \left\{ \frac{dr^2}{(1 - kr^2)} 
   + r^2 (d\theta^2 + \sin^2 \theta d\chi^2) \right\} 
             \label{gRWN}   \end{equation} 
 has $N$ $C^0$ and $N(t=0) = 0$ on the signature change surface 
$\Sigma$.  The $tt$ and $rr$ EFEs $G^\alpha_ \beta = \kappa 
T^\alpha_\beta$, evaluate to [24] 
 \begin{eqnarray} 
  G^t_t = - \frac{3}{a^2} \left\{ k + \frac{(a')^2}{N} \right\} & = & 
     - \frac{\kappa}{2} \left\{ \frac{(\phi')^2}{N} + V \right\} 
                       = \kappa T^t_t  \label{EFEttN}  \\ 
  G^r_r = \left\{ \frac{a' N'}{a N^2} - \frac{(2 a a'' + (a')^2)} 
    {a^2 N} - \frac{k}{a^2} \right\} & = & \frac{\kappa}{2} \left\{ 
    \frac{ (\phi')^2 }{ N } - V \right\} = \kappa T^r_r  
\label{EFErrN} 
 \end{eqnarray} 
  where ${}' = \partial / \partial t$, the $\theta\theta$ and 
$\chi\chi$ equations adding nothing new.  The covariant and 
contravariant forms re-scale the $tt$ but not the $rr$ equations by 
factors of $N$.  The 
 Klein-Gordon equation is implied by the EFEs. 

     Strictly speaking, these equations are not well defined on 
$\Sigma$ except in a limiting sense as $N \rightarrow 0$, and for $k$ 
and $V$ 
 non-zero, $G^{tt} = \kappa T^{tt}$ is divergent in this coordinate 
system, regardless of the behaviour of $a'$ and $\phi'$ near $N = 0$.  
Similarly, the conservation equation $\nabla_\alpha G^\alpha_\beta$ or 
$\nabla_\alpha G^{\alpha \beta}$ is badly defined at $N = 0$.  The 
latter form is especially problematic, since $G^{tt}$ is divergent for 
$k \neq 0$.  One {\it could} therefore question whether the EFEs and 
the conservation equation are well defined on $\Sigma$.  
 A reasonable possibility is to evaluate these equations away from $N 
= 0$ and rescale by factors of $N$, to obtain equations which {\it 
are} well defined at $N = 0$.  The `second order equations' of [12] 
and equations (2) of [14] are essentially 
 re-scalings by factors of $N$, of the Einstein, Raychaudhury, 
 Klein-Gordon, and/or Friedmann equations.  Note that the re-scaling 
itself is only well-defined in the limit as $t \rightarrow 0$, not at 
$t = 0$, itself.  
 \\[-2mm] 

 \noindent{\it The Darmois approach with real ``time"} 

     In this approach, we postulate that two metrics of different 
signature, each of which solves the respective constant signature form 
of the EFEs, may be pieced together so that the Darmois junction 
conditions are satisfied at the signature change surface $\Sigma$.  
 Even if the EFEs are well defined at $\Sigma$ in each solution, the 
resulting matter fields are not necessarily smooth when put together.  
This is true with or without signature change. 

     We patch together the Lorentzian and Euclidean signature forms of 
the above metric (\ref{gRWN}) ($M^-$ and $M^+$), identifying the two 
$t = 0$ surfaces in each to form the junction surface $\Sigma$.  Using 
the notation of [1], the strictly classical treatment requires a real 
proper ``time" 
 \begin{equation}  d\tau = \sqrt{-\epsilon N}\; dt 
  \;\;,\;\;\;\;\;\;\;\;\;  \epsilon = \left\{ 
   \begin{array}{llll} 
     +1, & t \leq 0, & N \leq 0 & \mbox{Euclidean signature metric} \\ 
     \\ 
     -1, & t \geq 0, & N \geq 0 & \mbox{Lorentzian signature metric} 
   \end{array}  \right. 
 \end{equation} 
 The choice of the Darmois approach, in which all quantities are well 
defined without 
 re-scaling, in the limiting sense given in [1], is made largely 
because of this extra measure of discontinuity, and partly because of 
the above-mentioned singularities in the metric and Einstein tensors.  
In $M^\pm$ we calculate 
 \begin{eqnarray} 
   {}^3\!\!g_{ij} & = & \mbox{diag}(g_{rr}, g_{\theta \theta}, 
     g_{\chi \chi}) = a^2 \mbox{diag} \left( \frac{1}{(1 - kr^2)}, 
        r^2, r^2 \sin^2 \theta \right)  \\ 
   n^\alpha & = & \left( \frac{1}{\sqrt{-\epsilon N}\;}, 0, 0, 0 
         \right) \;\;,\;\;\;\;\;  n_\alpha n^\alpha = \epsilon  \\ 
   l_\alpha & = & (\sqrt{-\epsilon N}\;, 0, 0, 0) \;\;,\;\;\;\;\; 
        l_\alpha l^\alpha = \epsilon  \\ 
   K_{ij} & = & \mbox{diag}(K_{rr}, K_{\theta \theta}, K_{\chi \chi}) 
     = \frac{a a'}{\sqrt{-\epsilon N}\;} \mbox{diag} \left( 
        \frac{1}{(1 - kr^2)}, r^2, r^2 \sin^2\theta \right)  \\ 
   K & = & \frac{3 a'}{a \sqrt{-\epsilon N}\;} 
          = 3 \left( \frac{d(\ln a)}{d\tau} \right)  \\ 
   {}^3\!R & = & \frac{6 k}{a^2}  \\ 
   {}^3\!\nabla_j K^{ij} & = & 0  \\ 
   {}^3\!\nabla_i K & = & 0  \\ 
   \tilde{T}^{00} & = & T^{\mu\nu} l_\mu l_\nu 
     = \epsilon \tilde{T}^0_0 
     = \frac{- \epsilon}{2} \left\{ \frac{(\phi')^2}{N} + V \right\} 
     = \frac{1}{2} \left\{ \left( \frac{d \phi}{d \tau} 
       \right)^2 - \epsilon V \right\}  \\ 
   \frac{a^2}{(1 - kr^2)} \tilde{T}^{01} & = & \tilde{T}^0_1 
    = T^\mu_\nu l_\mu e^\nu_1 = 0 
 \end{eqnarray} 
 where $n^\alpha$ and $l_\alpha$ are chosen to point from $M^-$ into 
$M^+$.  The tilde indicates a quantity evaluated in geodesic normal 
coordinates based on $\Sigma$, used for the purpose of easy 
interpretation, and for compact notation indicating invariants. With 
the notation 
 \begin{equation} 
   [Z] = \left. Z \right|^+_\Sigma 
         - \left. Z \right|^-_\Sigma \;\;,\;\;\;\;\;\;\;\; 
   \left. Z \right|_\Sigma = \mbox{Lim}_{\;\rightarrow \Sigma} \;\; Z 
 \end{equation} 
 for the jump in the value of $Z$ across $\Sigma$, and choosing to 
keep $r$, $\theta$ and $\phi$ continuous through $\Sigma$, the first 
Darmois conditions $[{}^3\!\!g_{ij}] = 0$ give the obvious relations 
 \begin{equation}  k^+ - k^- = [k] = 0 \;\;,\;\;\;\;\;\;\;\;\; 
a^+_\Sigma - a^-_\Sigma = [a] = 0   \label{Dcond1}   \end{equation} 
 and the second Darmois conditions $[K_{ij}] = 0$ give 
 \begin{equation} 
   \left. \frac{a'}{\sqrt{-\epsilon N}\;} \right|^+_\Sigma 
 - \left. \frac{a'}{\sqrt{-\epsilon N}\;} \right|^-_\Sigma 
 = \left[ \frac{a'}{\sqrt{-\epsilon N}\;} \right] 
 = 0 
 = \left. \frac{d a}{d\tau} \right|^+_\Sigma - 
   \left. \frac{d a}{d\tau} \right|^-_\Sigma 
 = \left[ \frac{d a}{d\tau} \right] 
         \label{Dcond2}    \end{equation} 
 where $\tau$ is the comoving proper ``time".  Appropriate 
supplementary junction conditions for the scalar field are 
 \begin{equation}  [\phi] = 0 \;\;,\;\;\;\;\;\;\;\;  
     [n^\alpha \partial_\alpha \phi]  
     = \left[ \frac{d \phi}{d \tau} \right] 
     = \left[ \frac{\phi'}{\sqrt{-\epsilon N}\;} \right] 
     = 0 \;\;,\;\;\;\;\;\;\;\; [V] = 0  \;\;,\;\;\;\;\;\;\;\; 
     \left[ \frac{dV}{d\phi} \right] = 0 
                 \label{phicond}  \end{equation} 
 From eqs (47)-(50) of [1], the various surface effects in the 
conservation law are then 
 \begin{eqnarray} 
     E(l_\alpha) = [\tilde{G}^{00}] = - {}^3 \!\!R 
        & = & - \frac{6 k}{a_\Sigma^2} 
        = \kappa [\tilde{T}^{00}] 
        = - \kappa V_\Sigma  \label{SEu00}  \\ 
     E(\overline{e}^1_\alpha) = [\tilde{G}^{01}] = 
        2({}^3\!\nabla_j K^{1j} - {}^3\!\!g^{1j} \; {}^3\!\nabla_j K) 
        & = & 0 
        = \kappa [\tilde{T}^{01}] = 0  \\ 
     E(n^\alpha) = [\tilde{G}^0_0] = K^2 - K_{ij} K^{ij} 
        & = & 6 \left( \frac{(a')^2}{-\epsilon N a^2 } \right)_\Sigma 
        = \kappa [\tilde{T}^0_0] 
        = \kappa \left( \frac{(\phi')^2}{-\epsilon N} \right)_\Sigma 
                     \label{SEm00}  \\ 
     E(\overline{e}_1^\alpha) = [\tilde{G}^0_1] = 
        2({}^3\!\nabla_j K_1^j - {}^3\!\nabla_1 K)  
        & = & 0 
        = \kappa [\tilde{T}^0_1] = 0 
 \end{eqnarray} 
 Assuming that $\Sigma$ is not already a singular surface in one or 
both of $M^-$ or $M^+$ separately, these quantities are all well 
defined in the limit as $N \rightarrow 0$ without re-scaling.  From 
the EFEs we obtain the restrictions 
 \begin{equation} 
  \sqrt{\frac{\kappa}{6}} \left( \frac{d \phi}{d \tau} \right)_\Sigma 
  = \left( \frac{d(\ln a)}{d \tau} \right)_\Sigma 
  \;\;,\;\;\;\;\;\;\;\;  \frac{\kappa V_\Sigma}{2} = 
     \frac{3 k}{a^2_\Sigma}   \label{SCcond}   \end{equation} 
 and while $(da/d\tau)_\Sigma = 0$ and $(d\phi/d\tau)_\Sigma = 0$ are 
possibilites, they are not demanded.  In any case it is not possible 
to remove the surface effect $E(l_\alpha) = - {}^3\!R|_\Sigma = - (6 
k/a_\Sigma^2)$ for $k \neq 0$, in this scenario. 

     In summary, we are given $a$ and $\phi$ of the form 
 \begin{equation} 
  a = a_+(t) \Theta(t) + a_-(t) \Theta(-t) \;\;,\;\;\;\;\;\;\;\; 
  \phi = \phi_+(t) \Theta(t) + \phi_-(t) \Theta(-t)   \label{apthet} 
 \end{equation} 
 such that $a_\pm$, $\phi_\pm$ $\in C^3$ separately solve the constant 
signature EFEs, as well as satisfying the Darmois and 
 Klein-Gordon field junction conditions (\ref{Dcond1})-(\ref{phicond}) 
and signature change condition (\ref{SCcond}).  
 We also assume 
 \begin{equation} 
   \lim_{t \rightarrow 0} \frac{a'_\pm}{\sqrt{-\epsilon N}\;} 
      \;\;,\;\;\;\;\; 
   \lim_{t \rightarrow 0} \frac{\phi'_\pm}{\sqrt{-\epsilon N}\;} 
      \;\;\;\;\; \mbox{and} \;\;\;\;\; 
   \lim_{t \rightarrow 0} \left( \frac{2 a''_\pm}{N} - 
   \frac{a'_\pm N'}{N^2} \right) 
      \;\;\;\;\; \mbox{all exist.} 
 \end{equation} 
 It follows from (\ref{Dcond1}) and (\ref{phicond}) that $[a] = 0 = 
[\phi]$, so 
 \begin{eqnarray} 
  a' & = & a'_+(t) \Theta(t) + a'_-(t) \Theta(-t) + [a] \delta(t) 
       = a'_+(t) \Theta(t) + a'_-(t) \Theta(-t)    \label{eq29}  \\ 
  \phi' & = & \phi'_+(t) \Theta(t) + \phi'_-(t) \Theta(-t) + 
                [\phi] \delta(t) 
          = \phi'_+(t) \Theta(t) + \phi'_-(t) \Theta(-t) \\ 
  a'' & = & a''_+(t) \Theta(t) + a''_-(t) \Theta(-t) + [a'] \delta(t) 
                           \label{eq32} 
 \end{eqnarray} 
 While condition (\ref{Dcond2}), $0 = [\dot{a}] =  [a']/\sqrt{|N|}\;$, 
certainly ensures $[a'] = 0$, we must verify that it is sufficient to 
eliminate distributions when we insert this into (\ref{EFEttN}) and 
(\ref{EFErrN}).  For example, putting $\dot{{}} \equiv d/d\tau$,
 \begin{eqnarray} 
    [\dot{a}] = 0 = [a] \;\;\;\;\;\; \Rightarrow \;\;\;\;\;\; 
        \ddot{a} &=& \ddot{a}_+ \Theta(t) + \ddot{a}_- \Theta(-t)  
           \nonumber  \\
  &=& 
 \left( -\frac{N' a'_+}{2 N^2} + \frac{a''_+}{N} \right) \Theta(t) + 
 \left( \frac{N' a'_-}{2 N^2} - \frac{a''_-}{N} \right) \Theta(-t) 
           \nonumber  \\
  \Rightarrow \epsilon \ddot{a} &=&
 \left( \frac{N' a'_+}{2 N^2} - \frac{a''_+}{N} \right) \Theta(t) + 
 \left( \frac{N' a'_-}{2 N^2} - \frac{a''_-}{N} \right) \Theta(-t) 
           \nonumber   \\
  &=& \left( \frac{N' a'}{2 N^2} - \frac{a''}{N} \right)   \label{ex} 
 \end{eqnarray} 
 and there is thus no distribution in $a''/N$.  Conversely, if 
$( (N' a')/(2 N^2) - a''/N )$ is free of distributions, then 
$\epsilon \ddot{a}$ is.  
 Therefore all terms in the EFEs are bounded, and {\it we have a 
solution} of (\ref{EFEttN}) and (\ref{EFErrN}), as well as the 
 Klein-Gordon equation.  We stress that the Darmois conditions are 
imposed on $a(t)$, $\phi(t)$ and $V(\phi)$ before calculating 
$G^\mu_\nu$ and $T^\mu_\nu$ from them.  Our assumptions are borne out 
by solutions to 
 (\ref{EFEttN})-(\ref{EFErrN}) which will be exhibited elsewhere [27].  
A rigorous treatment of distributions in the presence of signature 
change will appear in [26]. 
 \\[-2mm] 

 \noindent{\it The distributional EFEs} 

     In the distributional approach, we must postulate a form of the 
EFEs that makes sense when $N(t)$ becomes the the step function $- 
\Theta(-t) + \Theta(t) = \varepsilon(t)$ and $N'$ becomes $2\delta$, 
where $\Theta(t)$ is the Heaviside function and $\delta(t)$ is the 
Dirac delta function.  In particular products of distributions must be 
avoided.  Then $t = \tau$ and 
 $\varepsilon = -\epsilon$.  
 There is currently no known way to derive such equations from first 
principles. 

     {\it Postulate A.}  The 
 re-scaled EFEs, $N \times$(\ref{EFEttN}) and $N^2 
\times$(\ref{EFErrN}), are such a form {\it if} it is {\it assumed} 
that $a$, $a'$, $\phi$, $\phi'$, $V(\phi)$ (and $dV/d\phi$) are at 
least $C^0$, and $a''$ (and $\phi''$) may contain a step function, but 
no distribution.  It then follows [12,14] that there is a 
distributional term in $N^2 \times G^r_r$, but not in $N^2 \times 
\kappa T^r_r$, unless $a'_\Sigma = 0$.  Given these assumptions with 
this postulated form of the distributional EFEs, $a'_\Sigma = 0$ is 
required. 

     However, inspection of $a''$ in (\ref{31}) and in solutions to 
 (\ref{EFEttN})-(\ref{EFErrN}) [27], with $N \rightarrow \varepsilon$ 
and $N'\rightarrow 2\delta$, suggests that this assumption may be too 
strong, and makes it questionable that $a'' \rightarrow 
(d^2a/d\tau^2)$ without any extra distributional terms.  In other 
words, the distributional term that appears in $N^2 \times$(3) when 
$N' \rightarrow \delta$ is only present because  
 (\ref{EFEttN})-(\ref{EFErrN}) were derived and re-scaled using the 
differentiable structure of a continuous $N$, which is not appropriate 
for a step function $\varepsilon$.

     {\it Postulate B.}  Recognising that this substitution 
effectively replaces lapse ``time" with proper ``time", $t \rightarrow 
\tau$, which affects the functional form of $a$, we consider it safer 
to transform from $t$ to $\tau \ne t$, rather than substituting 
functions with distributions.  We write $\dot{{}} \equiv d/d\tau$, 
$\ddot{{}} \equiv d^2/d\tau^2$ and we continue to assume the Darmois 
and scalar field junction conditions, 
 (\ref{Dcond1})-(\ref{phicond}), to eliminate any distributions in the 
following transformations: 
 \begin{eqnarray} 
         \tau(t) & = & \left\{ \begin{array}{l} 
    \tau_+ = \int_0^t \sqrt{N(t)}\; dt \;,\;\;\; t \geq 0 \\ 
      \\ 
    \tau_- = -\int^0_t \sqrt{-N(t)}\; dt \;,\;\;\; t \leq 0 
        \end{array} \right.  \\ 
     a & = & a(\tau(t)) \;,\;\;\;\; 
         a' = \left\{ \begin{array}{l} 
    \sqrt{N}\; \dot{a} \;,\;\; t \geq 0 \\ 
      \\ 
    \sqrt{-N}\; \dot{a} \;,\;\; t \leq 0 \\ 
        \end{array} \right. \;,\;\;\;\; 
         a'' = \left\{ \begin{array}{l} 
     N \ddot{a} + \frac{N'}{2 \sqrt{N}\;} \dot{a} 
             \;,\;\; t \geq 0 \\ 
     \\ 
     - N \ddot{a} - \frac{N'}{2 \sqrt{-N}\;} \dot{a} 
             \;,\;\; t \leq 0 
        \end{array} \right.   \label{31}  \\ 
     \phi & = & \phi(\tau(t)) \;\;,\;\;\;\;\; 
     \phi' = \left\{ \begin{array}{l} 
    \sqrt{N}\; \dot{\phi} \;,\;\;\; t \geq 0 \\ 
      \\ 
    \sqrt{-N}\; \dot{\phi} \;,\;\;\; t \leq 0 \\ 
        \end{array} \right. 
 \end{eqnarray} 
 Using these expressions in (\ref{eq29})-(\ref{eq32}) and substituting 
the result into (\ref{EFEttN}) and (\ref{EFErrN}) leads to: 
 \begin{eqnarray} 
  G^t_t = - \frac{3}{a^2} \left\{ k - \epsilon (\dot{a})^2 \right\} 
   & = & 
     - \frac{\kappa}{2} \left\{ -\epsilon (\dot\phi)^2 + V \right\} 
                       = \kappa T^t_t  \label{EFEttTD}  \\ 
  G^r_r = \left\{ \epsilon \frac{(2 a \ddot{a} + (\dot{a})^2)} 
    {a^2 } - \frac{k}{a^2} \right\} & = & \frac{\kappa}{2} \left\{ 
    -\epsilon (\dot\phi)^2 - V \right\} = \kappa T^r_r  
           \label{EFErrTD} 
 \end{eqnarray} 
 where we have again used (\ref{ex}).  We note that this form of the 
equations does not contain $N$ or $N'$, nor does it need rescaling, so 
it seems more natural for a proper time solution.  We have {\it 
assumed} that $a$, $\dot{a}$, $\phi$, $\dot{\phi}$, $V(\phi)$, (and 
$dV/d\phi$) are at least $C^0$, so that $\ddot{a}$ (and $\ddot{\phi}$) 
may contain a step function but no distribution, and we postulate 
(\ref{EFEttTD}) and (\ref{EFErrTD}) for the field equations when $N 
\rightarrow \varepsilon$.  All of the solutions in [27] satisfy these 
EFEs.  In particular, $\dot{a}_\Sigma = 0$ is {\it not} required.  
 \\[-2mm] 

 \noindent{\it The Lagrangian approach with imaginary ``time"} 

    We postulate a Lagrangian approach in which the squared momenta 
are required to be smooth.  The action for this case, in terms of the 
gravitational and 
 Klein-Gordon Lagrangian densities can be written as 
 \begin{eqnarray} 
    S & = & \int ({\cal L}_G + 2 \kappa{\cal L}_{KG}) d^4x \\ 
    {\cal L}_G & = & \sqrt{-g}\; R 
     = \sqrt{{}^3\!\!g N}\; ({}^3\!R + K_{ij} K^{ij} - K^2) \\ 
    {\cal L}_{KG} & = & - \frac{\sqrt{-g}\;}{2} \; 
  \{(\partial_\mu \phi) g^{\mu \nu}(\partial_\nu \phi) + V(\phi)\} 
 \end{eqnarray} 
 where $R$ and $g$ are the Ricci scalar and the metric determinant, 
and ${}^3\!\!g = a^6 r^4 \sin^2 \theta/(1 - kr^2)$ is the determinant 
of the 
 3-metric ${}^3\!\!g_{ij}$ of (\ref{gRWN}).  Leaving the 
 4-volume element $\sqrt{-g}\; d^4x$ in this form leads to an {\it 
effective} imaginary proper ``time" 
 \begin{equation}   d\sigma = \sqrt{N}\; dt   \end{equation} 
 in the Euclidean region, $N < 0$.  This leads to an imaginary unit 
``normal'' $\hat{n}^\alpha = dx^\alpha/d\sigma = (1/\sqrt{N}, 0, 0, 
0)$ for $N<0$ which still obeys $\hat{n}^\alpha \hat{n}_\alpha = -1$.  
In this way this region is not Euclidean in the classical sense.  It 
also means the extrinsic curvature and the momenta change from 
imaginary to real across $N = 0$, as noted by Hayward [12], 
 \begin{eqnarray} 
   \psi & = & \frac{\partial {\cal L}}{\partial \phi'} = 
     \hat{n}^\alpha \partial_\alpha \phi 
    = \frac{\phi'a^3 r^2 \sin \theta}{\sqrt{N (1 - kr^2)}\;} 
                   \label{psiIm}  \\ 
   K_{ij} & = & \hat{n}^\alpha \nabla_\alpha {}^3\!\!g_{ij} 
     = \frac{{}^3\!\!g'_{ij}}{\sqrt{N}\;}  \\ 
   p^i_j & = & {}^3\!\!g_{jk} \frac{\partial {\cal L}}{\partial 
     {}^3\!\!g'_{ik}} = \sqrt{{}^3\!\!g}\; (K^i_j - K \delta^i_j) 
       \;\;,\;\;\;\;\; K = K^m_m  \\ 
         & = & - \frac{2 a^2 a' r^2 \sin \theta} 
                   {a \sqrt{N (1 - kr^2)}\;} \delta^i_j  \label{pijIm} 
 \end{eqnarray} and the squared momenta change sign, 
 \begin{eqnarray} 
   \psi^* & = & \psi^2 = 
     \frac{(\phi')^2 a^6 r^4 \sin^2 \theta}{N (1 - kr^2)}  \\ 
   (p^*)^i_k & = & p^i_j p^j_k 
    = K^i_j K^j_k - 2 K K^i_k + K^2 \delta^i_k  \\ 
  & = & \frac{4 (a')^2 a^4 r^4 \sin^2 \theta}{N (1 - kr^2)} \delta^i_k 
 \end{eqnarray} 
 implying $\psi^*$, $(p^*)^{ij}$, $\psi$, $p^{ij}$ and $K_{ij}$ must 
all be zero at the change surface {\it assuming} the momenta are 
required to be smooth and $a$, $\phi$, $\phi'$ and $a'$ all real. 
 \\[-2mm] 

 \noindent{\it The Lagrangian approach with real ``time"} 

     One could argue that it is more reasonable to define the action 
using a real volume element $\sqrt{|g|}\; d^4x$ 
 \begin{eqnarray} 
    S & = & \int ({\cal L}_G + 2 \kappa{\cal L}_{KG}) d^4x \\ 
    {\cal L}_G & = & \sqrt{|g|}\; R 
     = \sqrt{{}^3\!\!g |N|}\; ({}^3\!R + K_{ij} K^{ij} - K^2) \\ 
    {\cal L}_{KG} & = & - \frac{\sqrt{|g|}\;}{2} \; 
  \{(\partial_\mu \phi) g^{\mu \nu}(\partial_\nu \phi) + V(\phi)\} 
 \end{eqnarray} 
 and this form leads to a real proper ``time" everywhere 
 \begin{equation}   d\tau = \sqrt{|N|}\; dt   \end{equation} 
 so that $n^\alpha = dx^\alpha/d\tau = (1/\sqrt{|N|}\;, 0, 0, 0)$ is 
real and genuinely spacelike $n^\alpha \hat{n}_\alpha = +1$ in the 
Euclidean region.  The momenta now stay real 
 \begin{eqnarray} 
  \psi & = & \frac{\phi'a^3 r^2 \sin \theta}{\sqrt{|N| (1 - kr^2)}\;} 
                        \label{psiRe} \\ 
   K_{ij} & = & \frac{{}^3\!\!g'_{ij}}{\sqrt{|N|}\;}  \\ 
   p^i_j & = & - \frac{2 a^2 a' r^2 \sin \theta} 
              {a \sqrt{|N| (1 - kr^2)}\;} \delta^i_j  \label{pijRe} 
 \end{eqnarray} and the squared momenta don't change sign, 
 \begin{eqnarray} 
   \psi^* & = & 
     \frac{(\phi')^2 a^6 r^4 \sin^2 \theta}{|N| (1 - kr^2)}  \\ 
   (p^*)^i_k & = & 
  \frac{4 (a')^2 a^4 r^4 \sin^2 \theta}{|N| (1 - kr^2)} \delta^i_k 
 \end{eqnarray} 
 Thus the above requirements of smoothness and reality lead to the 
conditions that $(a')^2/|N|$ and $(\phi')^2/|N|$ be smooth, in 
agreement with the real ``time" Darmois based approach. 
 \\[-2mm] 

 \noindent{\it The Darmois approach with imaginary ``time"} 

     We now apply the Darmois based approach to the quantum cosmology 
inspired model of signature change with an imaginary proper ``time" in 
the Euclidean region, using the same metric.  We list quantities that 
are different from above. 
 \begin{eqnarray} 
  d\sigma & = & \sqrt{N}\; dt = \sqrt{-\epsilon}\; d\tau  \\ 
   \hat{n}^\alpha & = & \left( \frac{1}{\sqrt{N}\;}, 0, 0, 0 
     \right) \;\;,\;\;\;\;\;  \hat{n}_\alpha \hat{n}^\alpha = -1  \\ 
   \hat{l}_\alpha & = & (\sqrt{N}\;, 0, 0, 0) \;\;,\;\;\;\;\; 
        \hat{l}_\alpha \hat{l}^\alpha = -1  \\ 
   K_{ij} & = & \frac{a a'}{\sqrt{N}\;} \mbox{diag} \left( 
        \frac{1}{(1 - kr^2)}, r^2, r^2 \sin^2\theta \right)  \\ 
   K & = & \frac{3 a'}{a \sqrt{N}\;} 
          = 3 \left( \frac{d(\ln a)}{d\sigma} \right)  \\ 
   \tilde{\tilde{T}}^{00} = T^{\mu\nu} \hat{l}_\mu \hat{l}_\nu 
           & = & \frac{1}{2} \left\{ \frac{(\phi')^2}{N} + V 
     \right\} = \frac{1}{2} \left\{ \left( \frac{d \phi}{d \sigma} 
       \right)^2 + V \right\}  \\ 
   \tilde{\tilde{G}}^{00} = G^{\mu\nu} \hat{l}_\mu \hat{l}_\nu 
           & = & \frac{3}{a^2} \left\{ k + 
    \frac{(a')^2}{N} \right\}   = \frac{3}{a^2} \left\{ k + \left( 
       \frac{d a}{d \sigma} \right)^2 \right\} 
 \end{eqnarray} 
 where the double tilde means that the geodesic normal proper ``time" 
coordinate $\sigma$ is imaginary where $N < 0$.  Applying the Darmois 
junction conditions as before, and again requiring $a$, $\phi$, $a'$ 
and $\phi'$ to be real, now gives 
 \begin{equation}  [k] = 0 \;\;,\;\;\;\;\;\; [a] = 0 \;\;,\;\;\;\;\;\; 
     \left[ \frac{d a}{d\sigma} \right] = 0  \end{equation} 
 which leads to 
 \begin{equation}   k^+ = k^- \;\;,\;\;\;\;\;\;\;\;\; 
        a_\Sigma^+ = a^-_\Sigma \;\;,\;\;\;\;\;\;\;\;\; 
        \left. i \frac{d a}{d\tau} \right|^+_\Sigma = 
        \left. \frac{d a}{d\tau} \right|^-_\Sigma = 0 = 
        \left. \frac{a'}{\sqrt{N}\;} \right|_\Sigma  \end{equation} 
 by the usual argument [18].  The supplementary scalar field junction 
conditions 
 \begin{equation}  [\phi] = 0 \;\;,\;\;\;\;\;\;\;\;  
     [\hat{n}^\alpha \partial_\alpha \phi]  
     = \left[ \frac{d \phi}{d \sigma} \right] 
     = 0 \;\;,\;\;\;\;\;\;\;\; [V] = 0  \;\;,\;\;\;\;\;\;\;\; 
       \left[ \frac{dV}{d\phi} \right] = 0  \end{equation} 
 lead to 
 \begin{equation} 
        \phi^+_\Sigma = \phi^-_\Sigma \;\;,\;\;\;\;\; 
        V^+_\Sigma = V^-_\Sigma \;\;,\;\;\;\;\; 
        \left. \frac{dV}{d\phi} \right|^+_\Sigma = 
        \left. \frac{dV}{d\phi} \right|^-_\Sigma \;\;,\;\;\;\;\; 
        \left. i \frac{d \phi}{d\tau} \right|^+_\Sigma = 
        \left. \frac{d \phi}{d\tau} \right|^-_\Sigma = 0 = 
        \left . \frac{\phi'}{\sqrt{N}\;} \right|_\Sigma 
 \end{equation} 
 Clearly $\phi'$ and $a'$ must go to zero {\it faster} than 
$\sqrt{N}\;$.  These requirements, derived using the imaginary Darmois 
approach, are in agreement with those of the imaginary Lagrangian 
approach and with quantum cosmological type signature change.  This 
approach guarantees that $G_{\mu \nu}$ and $T_{\mu \nu}$ are smooth 
and finite, and conservation is observed, since the surface effects 
are all zero too 
 \begin{equation} 
   E(\hat{l}_\alpha) 
            = \frac{3}{a^2_\Sigma} \left[ \frac{(a')^2}{N} \right] 
            = \frac{1}{2} \left[ \frac{(\phi')^2}{N} \right] 
            = 0 \;\;,\;\;\;\;\; 
   E(\overline{e}^1_\alpha) = 0 \;\;,\;\;\;\;\; 
   E(\hat{n}^\alpha) = - E(\hat{l}_\alpha) = 0 \;\;,\;\;\;\;\; 
   E(\overline{e}_1^\alpha) = 0 
 \end{equation} 
 Of course $G^{tt}$ and $T^{tt}$ are still divergent. 
 \\[-2mm] 

 \noindent{\it The EFEs with real ``time"} 

     When the EFEs (2)-(3) are calculated from (1), it is not 
necessary to specify the real or imaginary nature of the ``time" 
direction, nor is it evident upon inspection.  This is because a 
smooth signature changing metric does not contain information on 
whether tangent vectors to geodesics in the ``time" direction change 
character across $\Sigma$. 

     In order to implement strictly classical signature change, we 
must include the fact that $n^\alpha$ changes from a (real) space-like 
direction to a (real) time-like direction; $n^\mu g_{\mu \nu} n^\nu = 
-\varepsilon$.  Taking the derivative of this along $n^\alpha$ gives 
[5] 
 \begin{equation} 
  n^\mu n^\nu n^\lambda \nabla_\lambda g_{\mu \nu} + 
  2 n^\mu g_{\mu \nu} a^\nu = 2 \delta(\tau) \;\;,\;\;\;\;\;\;\;\; 
  a^\nu = n^\lambda \nabla_\lambda n^\nu 
 \end{equation} 
 So on $\Sigma$, either $a^\nu$ is not orthogonal to $n^\nu$, or 
$\nabla_\lambda g_{\mu \nu}$ is not zero, and the latter seems more 
likely given that signature change is a metric phenomenon.  In any 
case, this introduces an extra degree of discontinuity not revealed by 
calculating the Einstein equations directly from the metric, so that 
we have to write down the results on either side of $\Sigma$, and then 
carefully examine the junction.  The fundamental role played by the 
unit normal was stressed in [2,6].  Also, because of the metric 
degeneracy, a classical relativist suspects the $t$ coordinate may be 
badly behaved, just as the Schwarzschild time $T$ is, in the exterior 
Schwarzschild metric at $R = 2M$.  
 Thus, with the EFEs (\ref{EFEttN})-(\ref{EFErrN}) calculated directly 
from the metric (1) we project into $n^\alpha$'s frame, obtaining 
 \begin{eqnarray} 
  \tilde{G}^0_0 = G^\alpha_\beta l_\alpha n^\beta = 
    \kappa T^\alpha_\beta l_\alpha n^\beta     \rightarrow 
  \frac{3}{a^2} \left\{ -k - \frac{(a')^2}{N} \right\} & = & 
     \frac{\kappa}{2} \left\{ -\frac{ (\phi')^2 }{N} - V \right\} 
                  \label{tilGm00}  \\ 
  \varepsilon^2 \tilde{G}_{00} = \tilde{G}^{00} = G^{\alpha \beta} 
     l_\alpha l_\beta = \kappa T^{\alpha \beta} l_\alpha l_\beta  
        \rightarrow \frac{3}{a^2} \left\{ (\varepsilon) k 
        + \frac{(\varepsilon)(a')^2}{N} \right\} & = & 
     \frac{\kappa}{2} \left\{ \frac{(\varepsilon)(\phi')^2}{N} 
         + (\varepsilon) V \right\} 
 \end{eqnarray} 
 Assuming that the values of $k$, $a$, $a'$, $N$, $N'$, $\phi'$, and 
$V$ are continuous through $\Sigma$ suggests the signature change 
conditions (\ref{SCcond}) 
 in order that the Einstein equations hold on both sides of $\Sigma$.  
However, going from Lorentzian ($\varepsilon = +1$, $N > 0$) to 
Euclidean ($\varepsilon = -1$, $N < 0$) regions, $\tilde{G}_{00}$, 
$\tilde{T}_{00}$, $\tilde{G}^{00}$, and $\tilde{T}^{00}$ clearly jump 
by $-(6 k / a^2_\Sigma) = - \kappa V_\Sigma$ 
 --- as in (\ref{SEu00}) --- and this jump cannot be removed for $k 
\neq 0$.  Since the quantities in (\ref{tilGm00}) are all finite 
(unless $\Sigma$ is already singular in $M^\pm$), and since the sign 
of $N$ changes across $\Sigma$, it follows that $\tilde{G}^0_0$, and 
$\tilde{T}^0_0$ also jump by $6 (d(\ln a) / d \tau)^2_\Sigma = \kappa 
(d \phi / d \tau)^2_\Sigma$ 
 --- as in (\ref{SEm00}).

     Similarly, projecting the momenta (\ref{psiIm}) and (\ref{pijIm}) 
into the ``time"-like frame defined by $n^\alpha$, gives us 
(\ref{psiRe}) and (\ref{pijRe}) again. 
 \\[3mm] 

 \noindent{\large \bf CONCLUSION} 

     We have just demonstrated that (a) the results of the strictly 
classical Darmois approach to signature change remain the same when 
squared-lapse coordinates are used; (b) when the Darmois approach is 
applied to the imaginary ``time" model of signature change, it leads 
to Hayward's results; (c) when the Lagrangian approach is applied to 
the strictly classical model of signature change, our results are 
recovered. 

     There is a basic conceptual problem with this imaginary ``time".  
Coordinate invariance and local Lorentz invariance are highly 
desireable features of classical Relativity.  Can they be shown to 
hold in general for a metric structure with positive definite 
signature that admits vectors with negative magnitudes, in some 
physically meaningful sense?  Are the Einstein equations well 
motivated?  How do we know what is `physically meaningful' in spaces 
with these properties?  These issues need scrutiny.  An interesting 
possible approach has been given in [22-23], in which the ``time" 
direction is complex, and the Wick angle is treated as a dynamical 
degree of freedom. 

     Clearly the difference in results is {\it entirely} due to a 
different philosophy of signature change, and {\it not at all} due to 
the different approaches to junction conditions, or to the choice of 
coordinates.  Strictly classical signature change does produce surface 
terms in the conservation law, but the imaginary time signature change 
has a Euclidean region with time-like features. 

     Hayward's approach focusses on maximising the smoothness of the 
covariant metric, the covariant Einstein tensor and the matter fields 
through $\Sigma$ in a particular type of coordinate system.  Our 
approach focusses on the smoothness of the geometry and what kind of 
jumps in physical measurables, i.e. scalar invariants, are allowed 
across $\Sigma$.  In his Lagrangian approach, Hayward effectively 
chooses imaginary ``time", and his method then obtains 
 well-defined re-scaled EFEs through the signature change, at the cost 
of 
 not-well-defined $G_{\mu \nu}$ and $T_{\mu \nu}$ and divergent 
$g^{\mu \nu}$, $G^{\mu \nu}$ and $T^{\mu \nu}$.  We (Hellaby and Dray) 
choose real ``time" and our method obtains 
 well-defined, finite, 
 coordinate-invariant expressions for the jumps in certain physical 
quantities at the signature change surface, at the cost of finite but 
discontinuous $g_{\mu \nu}$ and $g^{\mu \nu}$.  Even in the absence of 
surface terms, bounded jumps in $G_{\mu \nu}$ and $T_{\mu \nu}$ are 
standard at boundaries.  

     We leave the reader to weigh the two philosophies of signature 
change, and the problems associated with each. 

     The EFEs for a step function metric are badly defined.  There is 
(as yet) no unique prescription for obtaining well defined 
distributional field equations.  Hayward's postulated form does lead 
to his results, but requires a 
 re-scaling of the EFEs.  We postulate a form that doesn't require 
 re-scaling and leads to our results.  In the absence of a derivation 
of the distributional, signature-changing EFEs from first principles, 
one should be careful not to claim that a particular form of these 
equations is ``the'' EFEs.  Rather, one must investigate and compare 
the properties of alternative definitions.  We emphasize that in 
addition to being unrescaled, (\ref{EFEttTD}) and (\ref{EFErrTD}) are 
equivalent to (\ref{EFEttN}) and (\ref{EFErrN}) for continuous $N$, 
whereas the distributional form of $N^2 \times$(\ref{EFErrN}) contains 
an extra distributional term that is not removed by the associated 
coordinate tranformation. 

     We regard Hayward's approaches to signature change as reasonable 
and interesting, but not the only possibilities. 
 \\[3mm] 

 \noindent{\large \bf ACKNOWLEDGEMENTS} 

     We wish to thank George Ellis and David Coule for important 
comments.  CH would like to thank the FRD for a research grant.  TD 
was partially funded by NSF grant PHY-9208494.  
 \\[3mm] 

 \noindent{\large \bf REFERENCES} 

[1] C. Hellaby and T. Dray, Phys. Rev. D {\bf 49}, 5096-5104 (1994). 

[2] T. Dray and C. Hellaby, J. Math. Phys., to appear. 

[3] G.F.R. Ellis, A. Sumeruk, D. Coule, and C. Hellaby, Class. Q. 
   Grav. {\bf 9}, 1535-54 (1992). 

[4] G.F.R. Ellis, Gen. Rel. Grav. {\bf 24}, 1047-68 (1992). 

[5] G.F.R. Ellis and K. Piotrkowska, Int. J. Mod. Phys. D {\bf 3}, 49 
   (1994). 

[6] M. Carfora and G.F.R. Ellis,  ``The Geometry of Classical Change 
   of Signature'', SISSA report (unpublished), to appear in Int. J. 
   Mod. Phys. D 

[7] A. Sumeruk, C. Hellaby, and G.F.R. Ellis, University of Cape Town 
   report (unpublished). 

[8] T. Dray, C.A. Manogue, and R.W. Tucker, Gen. Rel. Grav. {\bf 23}, 
   967-71 (1991). 

[9] T. Dray, C.A. Manogue, and R.W. Tucker, Phys. Rev. D {\bf 48}, 
   2587-90 (1993). 

[10] T. Dray, C.A.Manogue, and R.W. Tucker, ``The Effect of 
   Signature Change on Scalar Field Propagation'', Oregon State 
   University report (Unpublished) (1993). 

[11] T. Dray, C.A.Manogue, and R.W. Tucker, ``Uniquness in the 
   Presence of Signature Change", Lancaster University report 
   (Unpublished) (1994). 

[12] S.A. Hayward, Class. Q. Grav. {\bf 9}, 1851-62 (1992); erratum 
   ibid {\bf 9} 2453 (1992). 

[13] S.A. Hayward, Class. Q. Grav. {\bf 10}, L7-11 (1993). 

[14] S.A. Hayward, Phys. Rev. D {\bf 52}, 7331-2 (1995).

[15] M. Kossowski and M. Kriele, Class. Q. Grav. {\bf 10}, 1157-64 
   (1993). 

[16] M. Kossowski and M. Kriele, Class. Q. Grav. {\bf 10}, 2363-71 
   (1993). 

[17] J.J. Halliwell and J.B. Hartle, Phys. Rev. D {\bf 41}, 1815-34 
   (1990). 

[18] G.W. Gibbons and J.B. Hartle, Phys. Rev. D {\bf 42}, 2458-68 
   (1990). 

[20] R. Kerner and J. Martin, Class. Q. Grav. {\bf 10}, 2111-22 
   (1993). 

[21] J. Martin, Phys. Rev. D {\bf 49}, 5086-95 
   (1994). 

[22] J. Greensite, Phys. Lett. B {\bf 300}, 34-7 (1993). 

[23] E. Elizalde, S.D. Odintsov, and A. Romeo, Class. Q. Grav. 
   {\bf 11}, L61-7 (1994). 

[24] Many results shown here were calculated using ``G.R.Tensor" by K. 
Lake and P. Musgrave (Physics Department, Queen's University, 
Kingston, Ontario, CANADA; E-mail: grtensor@astro.queensu.ca), running 
under Maple V release 2. 

[26] R.W. Tucker, T. Dray, D. Hartley, C.A. Manogue, and P. Tuckey, 
{\it Tensor Distributions in the Presence of Degenerate Metrics}, in 
preparation. 

[27] C.Hellaby {\it Solutions to Signature-Changing Field Equations}, 
in preparation. 

 \end{document}